\documentclass[useAMS,usenatbib]{mn2e}
\usepackage{epsfig}
\usepackage{graphicx}
\usepackage{amsmath}

\newcommand{\ba}{\begin{eqnarray}}
\newcommand{\ea}{\end{eqnarray}}
\newcommand{\be}{\begin{equation}}
\newcommand{\ee}{\end{equation}}

\def\go{\mathrel{\raise.3ex\hbox{$>$}\mkern-14mu
             \lower0.6ex\hbox{$\sim$}}}
\def\lo{\mathrel{\raise.3ex\hbox{$<$}\mkern-14mu
             \lower0.6ex\hbox{$\sim$}}}

\begin{document}

\title[Rossby Wave Instability in Magnetic Discs]
{Rossby Wave Instability in Accretion Discs with Large-Scale Poloidal
Magnetic Fields}
\author[C. Yu and D. Lai]
{Cong Yu$^{1,2,3}$\thanks{Email:cyu@ynao.ac.cn},
and Dong Lai$^2$\\
$^1$ Yunnan Astronomical Observatory, Chinese Academy of Sciences, Kunming, China \\
$^2$ Department of Astronomy, Cornell University, Ithaca, NY
14853, USA \\
$^3$ Key Laboratory for the Structure and Evolution of Celestial
Objects, Chinese Academy of Sciences, Kunming, 650011, China}

\pagerange{\pageref{firstpage}--\pageref{lastpage}} \pubyear{2011}

\label{firstpage}
\maketitle
\begin{abstract}
We study the effect of large-scale magnetic fields on the
non-axisymmetric Rossby wave instability (RWI) in accretion discs.
The instability develops around a density bump, which is likely
present in the transition region between the active zone and dead
zone of protoplanetary discs. Previous works suggest that the
vortices resulting from the RWI may facilitate planetesimal
formation and angular momentum transport.  We consider discs
threaded by a large-scale poloidal magnetic field, with a radial
field component at the disc surface.  Such field configurations
may lead to the production of magnetic winds or jets. In general,
the magnetic field can affect the RWI even when it is sub-thermal
(plasma $\beta\sim 10$). For infinitely thin discs, the
instability can be enhanced by about 10 percent. For discs with
finite thickness, with a radial gradient of the magnetic field
strength, the RWI growth rate can increase significantly (by a
factor of $\sim 2$) as the field approaches equipartition ($\beta
\sim 1$). Our result suggests that the RWI can continue to operate
in discs that produce magnetic winds.
\end{abstract}


\section{Introduction}

Radial density structures can be common in astrophysical discs.  For
instance, in proto-planetary discs, different degrees of coupling of
magnetic field with disc gas could lead to a transition region where
the disc material piles up (e.g., Gammie 1996;
Matsumura \& Pudritz 2003; Terquem 2008).
Proto-planets can also induce dips/gaps in the disc through tidal
interaction
(Lin \& Papaloizou 1986; Ward 1997). In accretion discs around
black holes or neutron stars, sharp density contrasts can be
created near inner disc edge inside which the disc gas plunges
inwards (e.g. Lovelace et al.~2009).

Such density structures lead to shear flows that are susceptible
to Kelvin-Helmholtz instability (Papaloizou \& Pringle 1985;
Papaloizou \& Lin 1989; Lithwick 2009).
A specific realization of this instability is the so-called Rossby
Wave Instability (RWI) (Lovelace et al.~1999; Li et
al.~2000,2001). For two-dimensional (vertically integrated)
barotropic discs, the RWI relies on the existence of an extremum
in the background fluid vortensity (also called potential
vorticity; see Drazin \& Reid 2004; Narayan et al.~1987), defined
by
$\zeta=\kappa^2/(2\Omega\Sigma)$,
where $\Sigma$ is the disc surface density,
$\Omega$ the disc rotation rate and
$\kappa^2=4\Omega^2 + 2r\Omega d\Omega/dr$
is the square of the radial epicyclic frequency.
Since Rossby waves propagate along the
gradient of vortensity, the instability can be understood
as arising from the interaction between two Rossby waves propagating on
each side of the vortensity extremum (e.g., Tagger 2001; Lai \& Tsang 2009).

The nonlinear development of the RWI in discs leads to the production
of vortices, which can accelerate the growth of meter-sized solids by
concentrating dust grains and thus enhance the formation of
planetesimals (Barge \& Sommeria 1995; Johansen et al.~2004; Varniere
\& Tagger 2006; Heng \& Kenyon 2010).
Such vortices may also induce sizable angular momentum transfer in
the radial direction (Li et al.~2001).  The vortices formed at the
edge of planetary gaps can affect the rate of planet migration (de
Val-Borro et al.~2007; Yu et al. 2010; Lin \& Papaloizou 2011a,b).


While most studies of the RWI deal with two-dimensional
(height-integrated) discs, Meheut et al.~(2010,2012) carried out the
global simulations of the RWI in 3D discs [see also
Lyra \& Mac Low (2012) for MHD simulations of 3D discs with no
vertical stratification]. They showed that the
Rossby vortices have significant vertical structure, which affects the
long-term evolution of the vortices. Meheut, Yu \& Lai (2012) and Lin
(2012) studied the linear RWI in 3D vertically stratified discs,
showing that while the RWI is basically a 2D instability,
appreciable vertical velocities can be induced in the disc.

Magnetic fields are present in most accretion discs and can
greatly change the dynamics of discs. It is generally thought that
turbulence induced by the magneto-rotational instability (MRI;
Balbus \& Hawley 1998) is responsible for angular moementum
transport in ionized discs. Recent simulations including ambipolar
diffusion, however, suggest that the MRI associated with
small-scale magnetic field may not be adequate to account for the
angular momentum transport in protoplanetary discs, even in the
``active zone''
(Bai \& Stone 2011).
On the other hand,
large-scale poloidal magnetic fields threading the accretion disc
can generate outflows/jets and play an important role in the
angular momentum transport (e.g., Blandford \& Payne 1982).
Large-scale magnetic fields are naturally present in the early phase
of star formation. They may be advected inward with the accretion flow
(e.g., Lubow et al.~1994), building up significant strength in the
disc. The large-scale magnetic field may
also counteract the disk self-gravity effect, thereby influencing
disc fragmentation (Lizano et al.~2010).


Recently, Yu \& Li (2009) showed from linear analysis
that a weak (sub-thermal) toroidal magnetic field in the disc
can completely surpress the RWI. This surpression results from
the modification by the magnetic field to the wave absorption
at the corotation resonance (see Fu \& Lai 2011).
In this paper, we study the effects of large-scale poloidal
magnetic fields on the RWI. As noted above, such a large-scale
field is essential for producing bipolar outflows and may be
necessary for angular momentum transport in the disc. Since the
growth of the RWI is primarily due to in-disc motion (Meheut et
al.~2012), we will focus on 2D disc dynamics in this paper.
However, we will consider the effect of finite disc thickness on
the RWI growth rate while neglecting vertical stratification.


Note that in our models (see Section 2), the large-scale poloidal
magnetic field provides a coupling between the dynamics of the
disc and the magnetosphere (assumed to be current-free). We
consider perturbations that have no vertical structure {\it
inside} the disc (i.e., the vertical wavenumber $k_z=0$). Thus, we
do not include MRI, which generally involves perturbations with
finite $k_z$.  Our basic disc-magnetosphere setup is
self-consistent. Similar setups have been considered by various
authors in different contexts (e.g., Spruit et al. 1995; Tagger \&
Pallet 1999; Lizano et al. 2010).

Our paper is organized as follows. In \S 2, the basic disc equations
with large-scale magnetic fields are derived. In \S 3, we present
the results of RWI of thin magnetized discs.
In \S 4, we consider the effects of finite disc thickness.
and we conclude in \S 5.

\section{Basic Equations}



We consider a thin conducting disc threaded by a magnetic
field.  The height-integrated mass continuity equation and momentum
equation read
\ba
&& \frac{\partial\Sigma}{\partial t}
+ \nabla_\perp\cdot\left( \Sigma \mathbf{u} \right) =0,\\
&& \frac{d \mathbf{u}}{dt} = -\frac{1}{\Sigma}\nabla_\perp P +
\frac{1}{4\pi\Sigma} B_z \left[ \mathbf{B} \right]^{+}_{-} +
\mathbf{g},
\label{momentumeqn} \ea where $\nabla_\perp$ is 2D operator
(acting on the disc plane), $\Sigma$, $P$ and ${\bf u}$ are the
surface density, height-integrated pressure and height-averaged
velocity, respectively, $\mathbf{g}=-g(r){\hat r}$ (with
$g=GM/r^2$) is the gravitational acceleration, $[{\bf
B}]^+_-\equiv {\bf B}(z=H)-{\bf B}(z=-H)$ (with $H$ the
half-thickness of the disc), and we have used $[B^2]^+_-=0$. In
this section we are considering infinitely thin discs, i.e., in
the limit $H/r \rightarrow 0$. As a result, we neglect the {\it
internal} magnetic force acting on the disc in Equation
(\ref{momentumeqn}).
The effect of finite disc thickness will be considered in Section
4.

For an equilibrium disc, the unperturbed velocity is
${\bf u}=(0,r\Omega,0)$ (in cylindrical
coordinates), with the angular velocity determined by
\begin{equation}
- \Omega^2 r = - \frac{1}{\Sigma}\frac{dP}{dr} - g +
\frac{B_z}{2\pi\Sigma}B_r^{+},
\label{eq:equil}
\end{equation}
where $B_r^+=B_r(z=H)=-B_r^-$. The unperturbed disc has
$B_\phi^+=0$. Note that $B_r$ is nonzero only outside the discs,
which has different signs just above and below the disc. Inside
the disc, $B_r$ is zero and the differential rotation of the disc
will not lead to generation of $B_{\phi}$ inside the disc.


Note that the disc dynamics are coupled with the large scale
poloidal magnetic field outside the disc, i.e., the disc
magnetosphere. Now consider small-amplitude perturbations of the
disc. Assuming that all perturbed quantities are proportional to
$\exp(im\phi-i\omega t)$, where $m=1,2,\cdots $ is the azimuthal
wave number and $\omega$ is the complex frequency, then the
linearized fluid perturbation equations read
%
\ba && - i \tilde{\omega}\ \delta\Sigma = -\nabla_\perp\cdot
\left(\Sigma\
\delta \mathbf{u}\right),\label{eq:dsigma}\\
&& - i \tilde{\omega}\ \delta u_r - 2\Omega\ \delta u_{\phi} =
-\frac{\partial}{\partial r}\ \delta h
+ \frac{B_z}{2\pi\Sigma} \ \delta B_r^+ \nonumber\\
&&\qquad\qquad\qquad\qquad\quad
+ \frac{B_r^+}{2\pi} \delta\left(\frac{B_z}{\Sigma} \right),\\
&&- i \tilde{\omega}\ \delta u_{\phi}+ \frac{\kappa^2}{2\Omega}\
\delta u_r = - \frac{i m}{r}\ \delta h +\frac{B_z}{2\pi\Sigma} \
\delta B_\phi^+,
\ea where $\tilde{\omega} = \omega - m\Omega$ is the
Doppler-shifted frequency, $\kappa$ is the radial epicyclic
frequency, and $\delta h=\delta P/\Sigma$ is the enthalpy
perturbation (we assume barotropic discs). The magnetic field
induction equation in the disc reads
\begin{equation}
- i \tilde{\omega}\ \delta B_z = -  \nabla_\perp \cdot \left( B_z\
\delta\mathbf{u}  \right).
\label{eq:dbz}\end{equation}
Combining Eqs.~(\ref{eq:dsigma}) and (\ref{eq:dbz}), we have
\be
\delta \left({B_z\over\Sigma}\right)=-\xi_r
{d\over dr}\left({B_z\over\Sigma}\right),
\label{eq:dbsigma}\ee
where $\xi_r=\delta u_r/(-i\tilde\omega)$
is the Lagrangian displacement. Equation (\ref{eq:dbsigma})
can also be derived from
$(d/dt)(B_z/\Sigma)=0$.
In terms of $\xi_r$ and $\delta h$, the magnetic field perturbation is
\be
\delta B_{z} = D_1 \xi_r + D_2 \delta h,
\label{eq:dbz0}\ee
with
\be
D_1=B_z{d\over dr}\left(\ln{\Sigma\over B_z}\right),
\quad \ D_2 = \frac{B_z}{c_s^2},
\ee
where $c_s$ is the sound speed.

%

To determine $\delta B_r^+$ and $\delta B_\phi^+$, we assume that
the magnetic field outside the disc is a potential field (e.g.,
Spruit et al.~1995; Tagger \& Pallet 1999).
This amounts to assuming that the Alfven speed above and below the
disk is sufficiently high that currents are dissipated quickly (on
the disc dynamical timescale). Define the magnetic potential
$\delta\Phi_M$ outside the disc via \be \delta{\bf B}=-{\rm
sign}(z)\nabla\delta\Phi_M. \ee Then $\delta\Phi_M$ satisfies the
Poisson equation (Tagger \& Pallet 1999) \be \nabla^2
\delta\Phi_{M} = -2\,\delta B_z\,\delta(z), \label{poisson} \ee
where $\delta(z)$ is the Dirac delta function. The integral
solution of (\ref{poisson}) is
\begin{equation}
\delta \Phi_{M}(r) = \int \delta B_z (r^{\prime})\,
\left[ \frac{\alpha}{2} \ b^{m}_{1/2}( \alpha ) \right]
dr^{\prime},
\label{lapcoeff1}
\end{equation}
where $\alpha = r^{\prime}/r$, and
the Laplace coefficient is defined by
\begin{equation}
b_s^{m}(\alpha) = \frac{2}{\pi} \int^{\pi}_0 \frac{\cos
m\phi}{(\alpha^2 + \epsilon_0^2 + 1 - 2 \alpha \cos\phi)^s}\
d\phi,
\end{equation}
with $\epsilon_0$ the softening parameter
(of order the dimensionless disk thickness $H/r$).
The perturbed radial magnetic field at the upper disc
surface is
\ba
&&\delta B_r^+ = - \frac{d}{dr}\delta \Phi_{M}\nonumber\\
&&\quad
=\int\!\delta B_z (r^{\prime})
\left(\frac{\alpha}{2r}\right)
\left[ b^{m}_{1/2}(\alpha)
+\alpha \frac{{\rm d}b^{m}_{1/2}(\alpha)}{{\rm d}\alpha}
\right] dr^{\prime},
\label{lapcoeff2}\ea
and the azimuthal field is
$\delta B_{\phi}^+ = -(im/r)\delta \Phi_{M}$.


In our numerical calculations, we use the variable
$\xi_r$ and $\delta h$. The corresponding perturbation
equations are
\ba
&&\frac{d \xi_r}{dr}=-\left(\frac{2 m \Omega}
{r\tilde{\omega}} + \frac{1}{r} + \frac{d\ln\Sigma}{dr} \right)\xi_r\nonumber\\
&&\qquad ~~ -\left(\frac{1}{c_s^2} - \frac{m^2}{r^2 \tilde{\omega}^2 } \right)
\delta h +
\frac{m^2}{r^2 \tilde{\omega}^2} \frac{B_z}{2\pi\Sigma}\delta\Phi_M,
\label{eq:dxir}\\
&&\frac{d}{dr}\delta h = \left( \tilde{\omega}^2-\kappa^2 +
\frac{B_r^+}{2\pi\Sigma}D_1
\right) \xi_r+ \frac{2 m \Omega}{r \tilde{\omega}}\delta h\nonumber\\
&&\qquad ~~~
- \frac{B_z}{2\pi\Sigma}\frac{d\delta\Phi_{M}}{dr} +
\frac{2 m \Omega}{r \tilde{\omega}} \frac{B_z}{2\pi\Sigma}
\delta\Phi_{M}.
\label{eq:dh}\ea

\section{Results for Thin Discs}

We consider a density bump in the disc of the form
\begin{equation}
\frac{\Sigma(r)}{\Sigma_0} = 1 + (A-1) \exp\left[
-\frac{(r-r_0)^2}{2\Delta^2}\right] \ .
\end{equation}
We adopt $A=1.2$ and $\Delta = 0.05 r_0$ throughout this paper. We
further assume
$c_s=0.1 r_0\Omega_0=$ constant.

\begin{figure}
 \includegraphics[width=84mm]{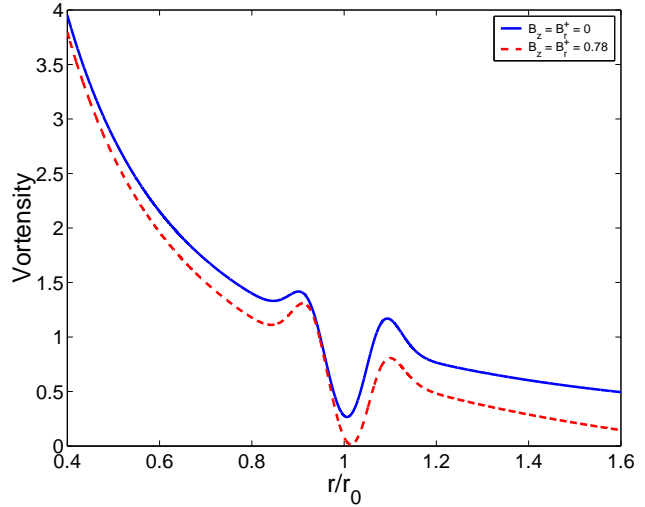}
 \caption{\label{fig2}
Vortensity profiles for magnetized discs with $\hat B_z=0.78$,
corresponding to the plasma $\beta\simeq 2$. The solid line is for the case
of $B_r^+=0$ and the dashed line for $B_r^+=B_z$.}
\end{figure}

In this section,  $B_z$ is taken as a constant independent of $r$
(we will consider varying $B_z$ in Sections 4-5).
The magnitude of $B_z$ is specified by the dimensionless ratio
\be
\hat B_z={B_z\over (\Sigma_0\Omega_0^2r_0)^{1/2}},
\ee
where the subscript ``0'' implies that the quantities evaluated at $r=r_0$.
The corresponding plasma $\beta$ parameter in the disc at $r=r_0$ is
\be
\beta = \frac{8\pi \rho c_s^2}{B_z^2}={4\pi H_0\over r_0}
\hat B_z^{-2},
\label{eq:beta}\ee
where we have used $H = c_s/\Omega_K\simeq c_s/\Omega$.
We solve for the equilibrium rotation profile using
Eq.~(\ref{eq:equil}).
In Fig.~\ref{fig2} we show two examples of the vortensity profiles
for the $B_r^+=0$ and $B_r^+=B_z$ cases.
Note that when $B_r^+ = 0$, the vortensity profile
is the same as a nonmagnetic disc. If $B_r^+ \neq 0$,
when $B_z$ increases, the epicyclic frequency approaches zero
gradually at the minimum, as does the vortensity.

Since the eigenfrequency $\omega$ is in general complex, equations
(\ref{eq:dxir})-(\ref{eq:dh}) are a pair of first-order differential
equations with complex coefficients which are functions of $r$. We
solve these equations using the relaxation method (Press et al.~1992),
replacing the ODEs by finite-difference equations on a mesh of points
covering the domain of interest (typically $0.4<r/r_0<1.6$).
According to Eqs. (\ref{eq:dbz0}), (\ref{lapcoeff1}), and
(\ref{lapcoeff2}), both the magnetic potential and its derivative
can be expressed in terms of linear combination of $\xi_r$ and
$\delta h$. For numerical convenience, we calculate the terms in
the square bracket in Eqs. (\ref{lapcoeff1}) and (\ref{lapcoeff2})
and store them for later use. Note that these terms are computed
only once and can be used repetitively. The wave equations
(\ref{eq:dxir}) and (\ref{eq:dh}) can be cast in a matrix form
that only deals with the variables $d\xi_r$ and $dh$. Standard
relaxation scheme can be applied to the resulting matrix.
We use uniform grid points in our calculations. The grid point
number is typically chosen to be 350. We implement the radiative
boundary conditions such that waves propagate away from the
density structure in both the inner and outer parts of the disk
(e.g., Yu \& Li 2009). The relaxation method requires an initial
trial solution that can be improved by the Newton$-$Raphson
scheme.
After iterations the initial trial solution converges to the
eigenfunction of the the two-point boundary eigenvalue problem.



\begin{figure}
 \includegraphics[width=84mm]{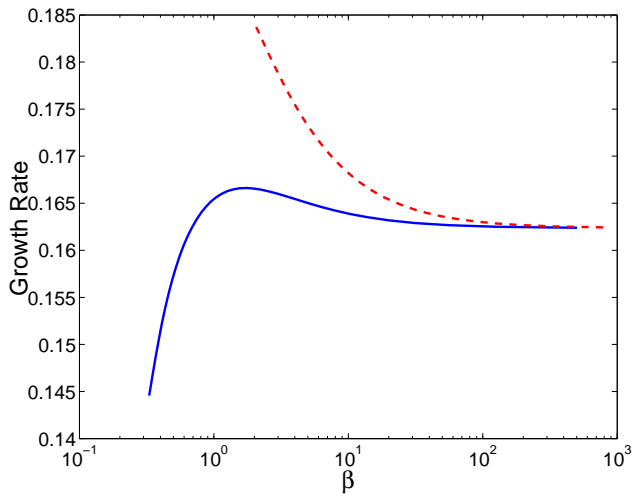}
 \caption{\label{fig1}
The growth rate
(in units of $\Omega_{K0}$)
of the $m=4$ RWI as a function of the plasma
$\beta$ parameter [with the corresponding vertical magnetic field
strength $B_z$
given by Eq.~(\ref{eq:beta})].
The solid line is the case of $B_r^+ = 0$, and
the dashed line $B_r^+=B_z$.}
\end{figure}

\begin{figure*}
   \vspace{1mm}
   \begin{center}
   \hspace{5mm}\psfig{figure=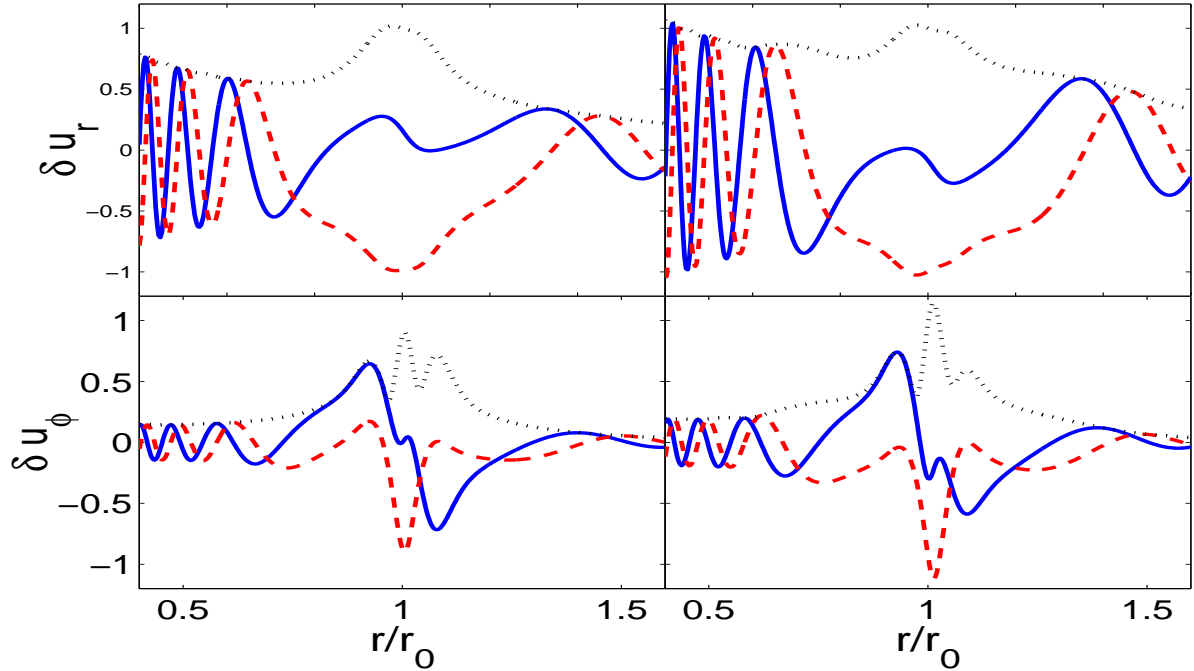,width=160mm,height=90mm,angle=0.0}
   \parbox{100mm}{{\vspace{6mm} }}
   \caption{
   \label{fig3a}
Eigenfunctions of the $m=4$ the Rossby modes trapped near the
density bump in the disc. The left panels show the non-magnetic
case, and the right panels have $B_z=B_r^+=0.78$. The top panels
show $\delta u_r$ (the solid line is for the real part, the dashed
line the imaginary part, and the dotted line the absolute value),
and the bottom panels show $\delta u_{\phi}$.}
   \end{center}
\end{figure*}

\begin{figure}
 \includegraphics[width=84mm]{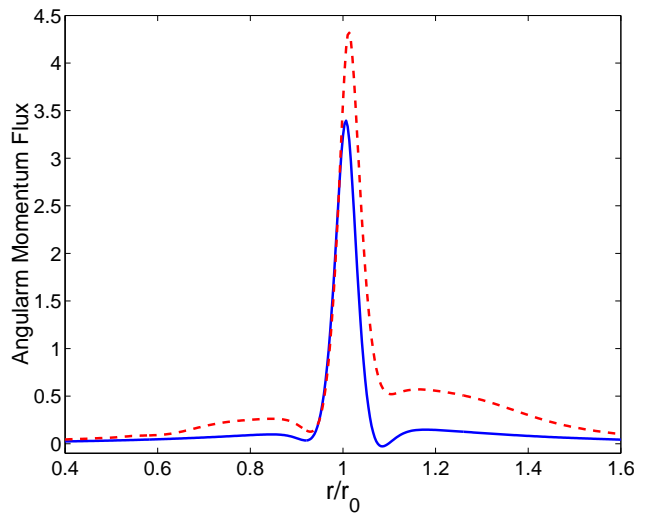}
 \caption{\label{fig3b}
The angular momentum flux associated with the Rossby mode. The
solid line is for the nonmagnetic case, and the dashed line for
${\hat B}_z={\hat B}_r^+=0.78$.
%
In both cases, the eigenfunction is normalized
so that the maximum $\delta u_r$ equals unity (near $r=r_0$). The
higher angular momentum flux at $r\sim r_0$ for the magnetic case
is consistent with the faster mode growth rate.
 }
\end{figure}

Figure 2 shows the growth rate of the RWI for different vertical
magnetic field strengths of the disc.
Real parts of the mode frequency for the solid line are all close
to $\omega_r \simeq 0.99 m\Omega_{K0}$. For the dashed line, real
parts of the mode frequency decrease from $\omega_r \simeq 0.99
m\Omega_{K0}$ to $\omega_r \simeq 0.95 m\Omega_{K0}$ with the
increase of magnetic field.
Since $B_z$ is assumed to
independent of $r$, the rotation profile of the disc is unaffected by
$B_z$ when $B_r^+=0$. In this case, we see that as $B_z$ increases,
the growth rate first increases slightly and then decreases. At
$\beta\sim 0.3$, the growth rate is reduced by more than $10\%$
compared to the $B_z=0$ value.  When $B_r^+=B_z$, the equilibrium
rotation profile of the disc is changed from the nonmagnetic disc.  In
this case, the growth rate increases monotonically with increasing
$B_z$, reaching about $10\%$ above the nonmagnetic value. Note that
the dashed line terminates at $\beta\simeq 2$, since an even stronger
magnetic field would make $\kappa^2$ negative.
Comparing the two cases depicted in Fig.~2, it appears that the
increase in the RWI growth rate mainly results from the deeper
minimum of the disc equilibrium vortensity profile induced by the
inclined ($B_r^+>0$) magnetic field.

Figure \ref{fig3a} gives some examples of the $m=4$ eigenfunctions
of Rossby modes trapped around the corotation radius, which is
also close to the density bump. The left panels show the
non-magnetized case, and the the right panels show the magnetized
case with $B_r^+ = B_z = 0.78$. The amplitude of eigenfunctions
are normalized so that the maximum absolute value of the radial
velocity perturbation $|\delta u_r|$ equals unity (this maximum
occurs at $r\simeq r_0$). The increase of $\delta u_r$ near the
inner boundary is due to the geometric effects (Meheut, Yu \& Lai
2012). A Rossby mode excited around $r_0$ by the density bump can radiate
into both the inner and outer parts of the disk as spiral
density waves. The relative phase shift between the real and
imaginary parts indicates this propagation.

To understand the origin of the enhanced RWI growth rate for
magnetized discs, we show in Fig.~\ref{fig3b} the angular momentum
flux associated with the eigenmode for the nonmagnetic disc and
for the disc with $B_z=B_r^+=0.78$.
The angular momentum flux $F(r)$ across a cylinder of radius $r$ is
given by (e.g., Goldreich \& Tremaine 1979)
\begin{equation}
F(r) =\left\langle r^2 \int^{2\pi}_0 \Sigma \delta u_r \delta
u_{\phi} d\phi \right\rangle = \pi \Sigma r^2 \Re \left( \delta
u_r \delta u_{\phi}^{*} \right)
\end{equation}
where $\langle\rangle$ designates time average and the
superscript $*$ denotes complex conjugate. Note that waves carry
negative (positive) angular momentum inside (outside) the
corotation. The net positive (outward) angular momentum flux
$F(r)$ around the corotation (close to the density bump) indicates
the growth of the RWI. Higher angular momentum fluxes imply
higher instability growth rates (see Fig. \ref{fig3b}).

We have attempted a variational-principle analysis to determine
the direct effect of magnetic force on the RWI (for a given
vortensity profile). Such an analysis did not yield simple,
unambiguous results, and therefore is not presented here. This is
consistent with the non-monotonic behavior exhibited by the solid
line in Fig.~2.

\section{Effects of Finite Disc Thickness}

\subsection{Model Equations}
In Sections 2-3, we have adopted the infinitely thin disc
approximation ($H/r\rightarrow 0$) and the "internal" disc
magnetic force is neglected. When finite thickness of the disc is
considered, an ``internal'' magnetic force term should be added to
the right-hand side of Eq.~(\ref{momentumeqn}): \be \mathbf{f} =
\frac{1}{\Sigma}\int\!dz\, \left[ - \nabla_\perp
\left(\frac{B^2}{8\pi} \right) + \frac{1}{4\pi} \left( \mathbf{B}
\cdot \nabla_\perp \right) \mathbf{B} \right]. \ee Obviously, to
include the 3D effect rigorously would require examining the
vertical stratification of the density and magnetic field inside
the disc -- this is beyond the scope of this paper [see Meheut et
al.~(2012) and Lin (2012) for the case without magnetic field].
Here we consider a simple model where the internal density of the
disc is assumed to be independent of $z$, so that \be \Sigma =
2\rho H,\quad \ P = 2 p H. \ee Then the internal magnetic force
simplifies to \be \mathbf{f} = \frac{1}{4\pi\rho}
\left[-\nabla_\perp \left(\frac{B^2}{2}\right) + \left(
\mathbf{B}\cdot \nabla_\perp\right)\mathbf{B}\right]. \ee

We assume that only vertical field exists inside the unperturbed disc.
The equilibrium rotational profile is then determined by
\begin{equation}\label{eq:equil3D}
- \Omega^2 r = - \frac{1}{\Sigma}\frac{dP}{dr} - g + %
\frac{B_z}{2\pi\Sigma}B_r^{+} %
- \frac{B_z}{2\pi\Sigma}\left(H \frac{d B_z}{dr} \right) \ . %
\end{equation}

To derive the modified perturbation equations including $\delta {\bf f}$,
it is convenient to define a new perturbation variable
$\delta\Pi$ in place of $\delta h$:
\begin{equation}
\delta \Pi \equiv \frac{\delta P}{\Sigma} + \frac{B_z\delta
B_z}{4\pi\rho} = c_s^2 \frac{\delta \rho}{\rho} + \frac{B_z\delta
B_z}{4\pi\rho}.
\end{equation}
After some algebra, the final disc perturbation equations
can be written in the following form:
\ba
&&\frac{d\xi_r}{dr}=-\left(\frac{2m\Omega}{r\tilde{\omega}} +
\frac{1}{r} + \frac{d\ln\Sigma}{dr} + D_4 \right) \xi_r \nonumber\\
&&\qquad\quad
+\left(\frac{m^2}{r^2\tilde{\omega}^2} - D_5\right)\delta\Pi
+ \frac{m^2}{r^2\tilde{\omega}^2}\frac{B_z}{2\pi\Sigma}\delta\Phi_{M},\\
&&\frac{d}{dr}\delta\Pi = \left(\tilde{\omega}^2 - \kappa^2
+\frac{B_r^+}{2\pi\Sigma}D_1
+ D_6 \right) \xi_r \nonumber\\
&&\quad\qquad + \left(\frac{2m\Omega}{r\tilde{\omega}}
+ D_7 - \frac{d\ln\rho}{dr}\right)\delta \Pi\nonumber\\
&&\qquad\quad +\frac{2m\Omega}{r\tilde{\omega}} \frac{B_z}{2\pi\Sigma}
\delta\Phi_{M} - \frac{B_z}{2\pi\Sigma}\,\frac{d\delta\Phi_{M}}{dr},
\ea
where
\ba
&& D_4 =\frac{-\ c_a^2}{c_s^2 + c_a^2}
{d\over dr}\left(\ln {\Sigma\over B_z}\right),\quad
c_a^2 = \frac{B_z^2}{4\pi\rho},\\
&&
D_5 = \frac{1}{c_s^2 + c_a^2},\quad  D_6 = \left[
\frac{1}{\rho}\frac{d(p+\frac{1}{8\pi} B_z^2)}{dr} \right] D_4,\\
&&D_7 = \left[ \frac{1}{\rho}\frac{d(p+\frac{1}{8\pi} B_z^2)}{dr}
\right] D_5.
\ea

\subsection{Results}

\begin{figure}
 \includegraphics[width=84mm]{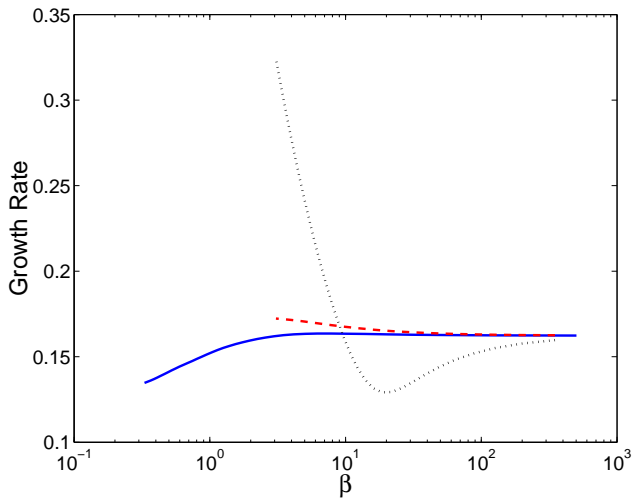}
 \caption{\label{fig4}
Growth rate of the
$m=4$
RWI as a function of the plasma $\beta$ parameter for discs with
finite thickness. The solid line is for the case of constant $B_z$
with $B_r^+=0$, while the dashed line for the case of constant
$B_r^+ = B_z$. Note that in these two cases, the growth rates are
very close to the cases depicted in Fig.~\ref{fig1} (where the
internal magnetic force $\delta{\bf f}$ is neglected). The dotted
line shows the case where $B_z(r)$ is given by
Eq.~(\ref{eq:Bzprofile}) with $B_{\mathrm{amp}} = -0.6
B_{\mathrm{mean}}$
and $B_r^{+} = B_z$.
As $B_{\rm mean}$ increases (or $\beta$ decreases), the growth rate first
decreases and then increases.
 }
\end{figure}

Finite disc thickness has two effects on the
RWI: one is the change in the equilibrium rotation profile
and the other is the direct effect of $\delta \mathbf{f}$.
We first examine the second effect by considering the
case of constant $B_z$ (no dependence of $r$).
Figure \ref{fig4} shows the result.
We find that for the parameter considered ($c_s=0.1$, or
$H/r=0.1$), the RWI growth rate is only slightly decreased (by
less than $5\%$) compared to the $H/r\rightarrow 0$ limit.

\begin{figure}
 \includegraphics[width=84mm]{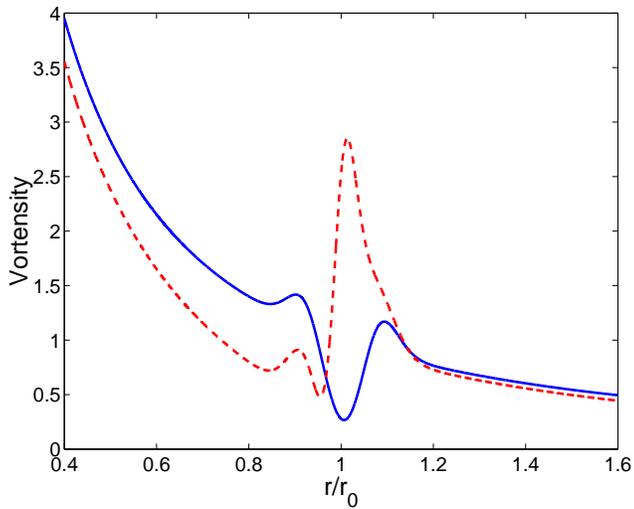}
 \caption{\label{fig7}
 Vortensity in both non-magnetized (blue curve)
 and strongly magnetized (red curve,
 $B_{\mathrm{mean}} = 0.64$, $B_{\mathrm{amp}} = - 0.6B_{\mathrm{mean}}$)
 situations are shown. The magnetic field gradient change the
 vortensity profile significantly.
 }
\end{figure}

Now consider the effect of the magnetic field gradient.
With finite disc thickness, such a gradient modifies the equilibrium
rotation profile. We use the following $B_z$ profile:
\begin{equation}
B_z = B_\mathrm{mean} + B_\mathrm{amp} \tanh\left(
\frac{r-r_0}{\Delta}\right). \label{eq:Bzprofile}\end{equation}
Thus, the vertical magnetic field changes from $B_{\rm
mean}-B_{\rm amp}$ at
$(r-r_0)/\Delta\ll -1$
to $B_{\rm mean}+B_{\rm amp}$ at $(r-r_0)/\Delta\gg 1$. The length scale of
the $B_z$ variation is the same as the size of the density bump.
The plasma $\beta$ defined in Eq.~(\ref{eq:beta}) is evaluated at
$r=r_0$ (where $B_z=B_{\rm mean}$).
We assume $B_r+=B_z$.
Figure \ref{fig7} shows some examples of the vortensity profiles.
The blue curve is for the non-magnetized case, the dashed line is
for the magnetized case where $B_{\mathrm{mean}} = 0.64$,
$B_{\mathrm{amp}} = - 0.6 B_{\mathrm{mean}}$. The magnetic field
gradient in Eq. (\ref{eq:equil3D}) alters the vortensity profile
dramatically. The dashed line profile corresponds to the highest
growth rate (dotted line, where $\beta \sim 3$) in Fig.
\ref{fig4}.

The dotted line of
Figure \ref{fig4} shows the RWI growth rate as a function of
$\beta$ in the case of $B_{\rm amp}=-0.6B_{\rm mean}$.
We see that the growth rate depends on $\beta$ in a non-monotonic
way: As $\beta$ decreases ($B_{\rm mean}$ increases),
the growth rate first decreases, and then increases, reaching
a factor of two compared to the non-magnetic value
(e.g., at $\beta\simeq 3$, corresponding to the vortensity profile
depicted by the dashed line of Fig.~6, the mode frequency is
$\omega_r = 1.02 m \Omega_{K0}$ with $m=4$, and the growth rate is
$\omega_i = 0.32 \Omega_{K0}$).

\begin{figure}
 \includegraphics[width=84mm]{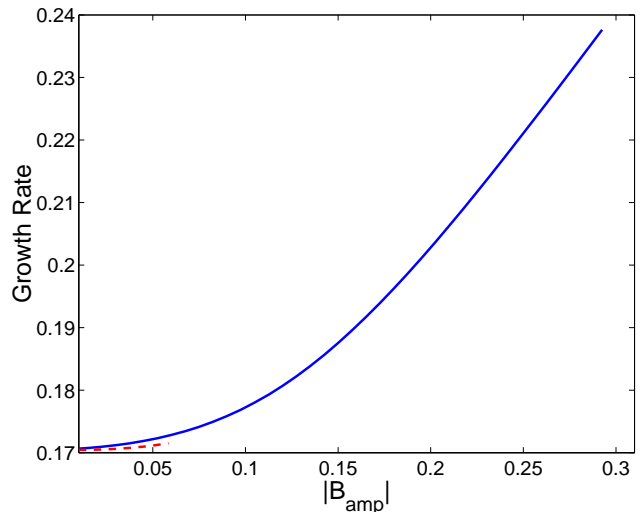}
 \caption{\label{fig5}
The dependence of the growth rate of the
$m=4$
RWI on on the $B_z$
gradient. The $B_z$ profile is given by Eq.~(\ref{eq:Bzprofile}),
with $B_{\rm mean}=0.5$ (corresponding to $\beta=5$) and $B_r^{+}
= B_z$. The solid line is for $B_{\rm amp}<0$ while the dashed
line for $B_{\rm amp}>0$. In such a strong magnetic field regime,
the magnetic field gradient (regardless of its sign) increases the
instability growth rate. Note that the lines terminate at the
point where the equilibrium rotation profile has $\kappa^2=0$ at
some radius near the density bump. }
\end{figure}

\begin{figure}
 \includegraphics[width=84mm]{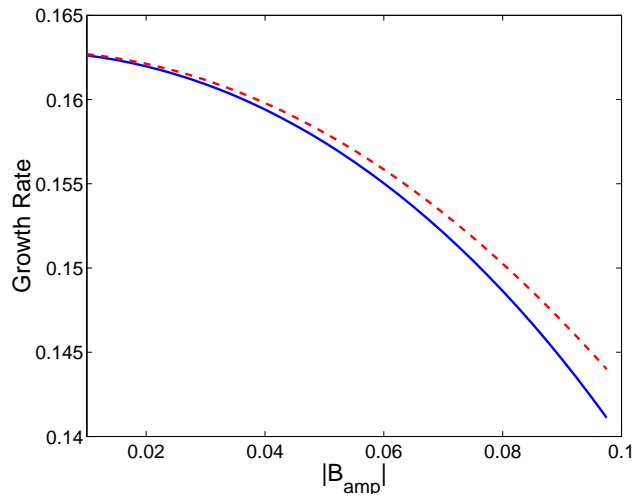}
 \caption{\label{fig6}
Same as Fig.~\ref{fig5} except for
${\hat B}_{\mathrm{mean}} = 0.1$
%
(corresponding to $\beta\sim120$). In such a weak
 magnetic field regime, the growth rate of the RWI tends to be
reduced by the magnetic field gradient.
 }
\end{figure}

Figure \ref{fig5} and Figure \ref{fig6}
illustrate how the RWI growth rate depends on the $B_z$ gradient,
for two different values of $B_{\rm mean}$.
For small $B_{\rm mean}$, the growth rate is reduced by the
$B_z$ gradient (regardless of the sign of he gradient); for
larger $B_{\rm mean}$, the growth rate is enhanced by the
$B_z$ gradient.


\section{Discussion and Conclusion}


In this paper we have carried out linear analysis of the Rossby
wave instability (RWI) in accretion discs threaded by large-scale
magnetic fields. The instability can develop around the extremum
of disc vortensity [$\zeta=\kappa^2/(2\Omega\Sigma)$, vorticity
divided by surface density], and may play an important role in
planetesimal formation and angular momentum transport in
proto-planetary discs, and may also generate variabilities in
black-hole accretion discs.  Our results show that the large-scale
magnetic field can affect (increase or decrease, depending on the
field configuration) the RWI growth rate even when it has a
sub-thermal strength (plasma $\beta\sim 10$ in the disc).  For
thin discs, the instability growth rate can be enhanced by the
magnetic field by up to $10\%$. We have also considered discs with
finite thickness (but ignoring vertical stratofication), and shown
that the RWI growth rate can be further increased (by a factor of
$\sim 2$ when $\beta\sim 1$) with a steep radial gradient in the
magnetic field strength. Just like the density bump, such field
gradient may be present around the transition region between the
active and dead zones in proto-planetary discs.

In general, the large-scale magnetic field influences the
RWI in two ways, either through the direct effect of
magnetic force on the perturbed fluid, or by modifying the
equilibrium disc vortensity profile. We find that the first
effect is typically smaller than the second.

Previous studies have shown that an ordered toroidal magnetic
field {\it inside} the disc can suppress the RWI, even when the
field strength is sub-thermal (Yu \& Li 2009). This comes about
because the in-disc toroidal field changes the behavior of wave
absorption at the corotation resonance (see Fu \& Lai 2011). By
contrast, the large-scale poloidal field considered in this paper
does not qualitatively affect the corotation resonance, and in
most cases leads to an enhanced RWI. Our calculations of discs
with finite thickness showed that the magnetic field gradient in
the radial direction can significantly affect the dynamics of RWI.
More rigorous treatments, taking into account the vertical
stratification of the disc and including both poloidal and
toroidal magnetic field, would be useful.

The magnetic field - disc configurations studied in this paper are
susceptible to the production of magneto-centrifugal winds. Recent
works suggested that in proto-planetary discs, MRI-induced
turbulence may be inadequate for transporting angular momentum and
magnetic winds may be necessary (Bai \& Stone 2011). Our result in
this paper implies that the RWI can operate under the
wind-producing conditions, potentially contributing to angular
momentum transport in the disc.


\section*{Acknowledgments}
This work has been supported in part by NSF grant AST-1008245,
NASA grants NNX12AF85G and NNX10AP19G. C. Y. thanks
the support from National Natural Science Foundation of China
(Grants 10703012, 11173057 and 11033008) and Western Light Young
Scholar Program of CAS. Part of the computation is performed at
HPC Center, Kunming Institute of Botany, CAS, China.


\end{document}